# Illusory Sense of Human Touch from a Warm and Soft Artificial Hand

John-John Cabibihan, Deepak Joshi, Yeshwin Mysore Srinivasa, Mark Aaron Chan, and Arrchana Muruganantham

JJ Cabibihan is with the Mechanical and Industrial Engineering Department, Qatar University, Doha, 2713, Qatar (email: john.cabibihan@qu.edu.qa; website: www.johncabibihan.com).

D Joshi is with the Electrical and Electronics Engineering Department, Graphic Era University, Dehradun, 248002, India.

YM Srinivasa and A Muruganantham are with the Electrical and Computer Engineering Department, National University of Singapore, Singapore, 117576, Singapore

MA Chan is with GE Global Research, Munich, 85748, Germany.

*For the video of the experiments, please click the link: http://youtu.be/lATSgG7CuQU*

# Abstract

To touch and be touched are vital to human development, well-being, and relationships. However, to those who have lost their arms and hands due to accident or war, touching becomes a serious concern that often leads to psychosocial issues and social stigma. In this paper, we demonstrate that the touch from a warm and soft rubber hand can be perceived by another person as if the touch were coming from a human hand. We describe a three-step process toward this goal. First, we made participants select artificial skin samples according to their preferred warmth and softness characteristics. At room temperature, the preferred warmth was found to be 28.4℃ at the skin surface of a soft silicone rubber material that has a Shore durometer value of 30 at the OO scale. Second, we developed a process to create a rubber hand replica of a human hand. To compare the skin softness of a human hand and artificial hands, a robotic indenter was employed to produce a softness map by recording the displacement data when constant indentation force of 1 N was applied to 780 data points on the palmar side of the hand. Results showed that an artificial hand with skeletal structure is as soft as a human hand. Lastly, the participants' arms were touched with human and artificial hands, but they were prevented to see the hand that touched them. Receiver operating characteristic curve analysis suggests that a warm and soft artificial hand can create an illusion that the touch is from a human hand. These findings open the possibilities for prosthetic and robotic hands that are lifelike and are more socially acceptable.

# I. INTRODUCTION

THE human hand has a complex anatomy consisting of bones, muscles, tendons, ligaments, arteries, nerves, and the protective layer of the skin [1, 2] (Figure 1). The bones form the innermost structure of the hand. The skin forms the outermost protective layer and acts as the primary interface to the external world [3]. It is comprised of several internal layers broadly classified as the stratum corneum, epidermis, dermis, and the hypodermis. Apart from the papillary ridges, sweat glands, and the blood vessels, the volar side of the hand is also densely packed with receptors to facilitate efficient tactile sensing [4]. A layer of subcutaneous fat pads lies in the palm and the digits to create a cushioning effect on the volar side of the hand.

The skin tissue of the human hand has versatile characteristics: it easily deforms with slight contact and stiffens when the contact force becomes large; it conforms to various shapes of objects, but returns to its original shape in seconds; it is tough against external elements, yet is soft enough to provide a comforting touch to a baby; it is warm enough to soothe someone else's physical and emotional aches and pains. Through touch, distinct emotions such as anger, fear, disgust, love, gratitude and sympathy can be communicated to others [5]. Given these remarkable characteristics, it is reasonable to mimic some of those features for prosthetic devices for those who lost hands or fingers due to an accident or war.

To date, significant advances have been made in the recovery of the motor and sensory functions that were lost due to amputation. With a surgical technique called targeted muscle reinnervation, Kuiken *et al* [6] demonstrated that real-time control of multiple degree of freedom joints for prosthetic arms and hands can be achieved when the nerves from the residual arm are transferred to alternative muscles sites and electromyography signals are recorded by electrodes at the skin surface.

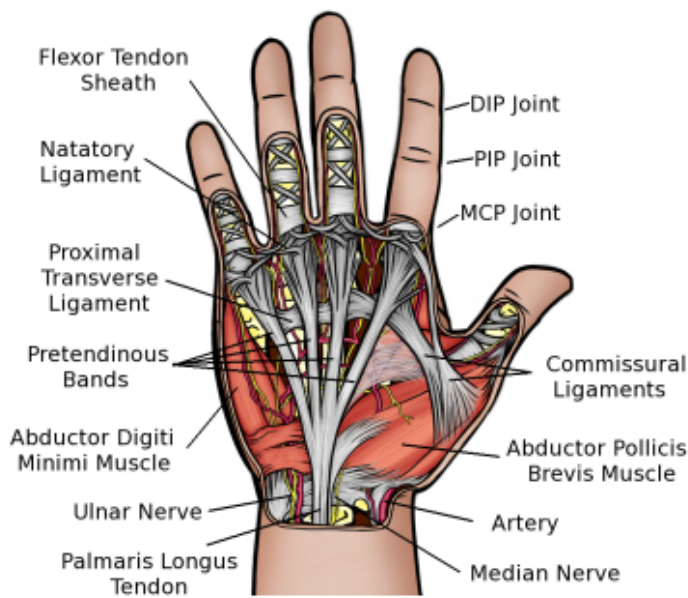

Fig. 1. Anatomy of the human hand. Depiction of the major tendons, ligaments, muscles, artery, and nerves at the volar side of the right hand.

Tactile feedback was found to be essential in order to control a prosthetic hand in an intuitive manner. Marasco *et al* [7] created an artificial sense of touch by coupling a pressure sensor on a prosthetic hand while feedback was achieved through a haptic interface (i.e. tactor), where proportional pressure is applied to stimulate the cutaneous nerves that were redirected to the skin of the residual limb. Self-reported and physiological measures from amputees suggest that a vivid sense of ownership of a prosthetic hand can be created by providing cutaneous tactile feedback. More recently, Raspapovic *et al* [8] demonstrated that an amputee's motor commands can be simultaneously decoded and sensory feedback can be delivered in real time for an amputee to bidirectionally control a prosthetic hand. From the contact information provided by the tactile sensors at the prosthetic hand, sensory information can be provided to the amputee by stimulating the median and ulnar nerve fascicles using implanted electrodes. For the first time, they showed that a blindfolded amputee can identify stiffness and shape of three different objects only from the tactile information provided by the prosthetic hand.

Upper limb loss has been found to dramatically change a person's body image [9-11]. To this end, creating a sensation that a foreign object is a part of the body has been investigated in the so-called Rubber Hand Illusion [12]. This illusion is created by applying synchronized brush strokes to a rubber hand in full view of the participant and to the participant's own hand, which is hidden behind a screen or under a table. Since its discovery, several reports have confirmed that an illusion of touch has been felt by non-amputees [13-15] and amputees [7, 16] alike, with both experimental groups experiencing ownership of the rubber hand. It was explained that the illusion occurs due to the attempt of brain's perceptual systems to interpret visual, tactile, and proprioceptive information resulting into a re-calibration of the location of touch and the felt position of the hand with the result that the touch appears to be felt by the rubber hand [12, 16].

Furthermore, it has been reported that the experience of body ownership applies only to objects that have the same appearance of the body part [17]. It was demonstrated that participants experienced a sense of ownership only for a realistic prosthetic hand and not for a plain wooden block or even a wooden hand. The study suggests that the object being viewed by the participants must fit with a reference model of the body in order to maintain a coherent sense that the object can be a part of the body. Considerable advances have been achieved to make prosthetic hands and fingers indistinguishable from the missing body parts not only in terms of anatomical structure [18, 19] but also in the replication of skin tone, pores, and hair [20, 21].

In addition to functional and aesthetic considerations, Murray [22] argued that the usage of prosthesis plays a social role in the lives of amputees—prosthesis use can ward off social stigmatization. This finding was corroborated by Ritchie *et al* [23] where they found that a prosthesis can help upper limb amputees cope, to feel normal again, and "not to stand out". Limb loss and prosthesis usage can alter a person's social life and the quality of social interactions [24-26].

All these findings are of particular importance as they provide evidence on the possibility of satisfying the functional, body image, aesthetic, and social requirements of those who lost a hand. What remains unknown, however, is whether an artificial hand would feel realistic to the person being touched.

In the present article, we consider a scenario where the user of a prosthetic hand touches another person. We ask whether experimental participants could feel that the touch from an artificial hand with warm and soft characteristics feels like touch coming from a human hand. We describe a three-step process to investigate this. First, we present the selection process for warm and soft synthetic skins that participants preferred to have on an artificial hand. We then show how we included these features into a replica of a human hand. Finally, we describe an experiment wherein participants were touched with a human hand and artificial hands, but they were prevented from seeing the hand that touched them.

## II. Perception Experiments with Artificial Skin Samples

### A. Participants

A total of 165 healthy subjects (95 males, 70 females, all 17-28 years old) were recruited from the National University of Singapore. Participation was voluntary.

### B. Experimental design

To determine the desired thermal and mechanical characteristics of an artificial hand, we asked participants to touch various skin samples and select a sample that they felt similar to human skin tissue. Participants were free to choose their own exploration strategy. They were not limited on the time to make a decision.

The skin samples were arranged to have a gradiation of soft and warm features and were laid out in a 4×4 array (Figure 2a). We designed each row to have a temperature gradient of 5°C at the sample's surface (i.e. 22, 27, 32 and 37°C). We wanted to determine the temperatures because we do not know which temperature is preferred by the subjects for a lifelike prosthetic hand. This is especially important because the room temperature has an effect on the skin temperature of the participant's hand [27]. For the columns, we selected 4 different materials of increasing softness. The details of the materials selected are provided in Table I. The lower Shore durometer value corresponds to a softer material.

TABLE I. DETAILS OF THE MATERIALS SELECTED

| Material Name | Manufacturer | Shore OO Durometer Number | Material Designation |
|---|---|---|---|
| Ecoflex, OO-30 | Smooth-On, USA | 30 | A |
| Prochima, GLS 40 | Prochima, Italy | 55 | B |
| Dragon Skin, series 20 | Smooth-On, USA | 70 | C |
| Polydimethylsiloxane (PDMS), Sylgard 184 | Dow Corning, USA | 86 | D |

The materials were selected based on their usage in earlier works on prosthetic or robotic skins [28, 29], embedding materials for tactile sensors [30, 31] and for soft robotics [32, 33]. All the skin samples were fabricated using standard standard moulding techniques. They were identically colored to remove any visual bias. The samples were colored using a flesh-toned pigment (Silc Pig, Smooth-On, USA). The pigment was mixed with silicone paint base (Psycho Paint, Smooth-On, USA) for the pigment to adhere properly.

### C. Design of the Embedded Heating

Figure 2b shows the assembly of a skin sample with a temperature sensor at the surface and a heating element at the base. The polyimide heater (Kapton heaters, Minco, USA), with dimensions of 6.35×25.4×0.25 mm³, has 7.1 Ω resistance with maximum current rating of 3 A. The 16 samples were glued on extruded polystyrene foam (Styrofoam). This material thermally insulated the samples from one another due to its low thermal conductivity of 0.03 W/m·K. All the samples were given 3.9 W of power. The ambient room temperature was maintained at 21°C, as measured with a thermometer in free air. The relative humidity was recorded to be about 50-60%.

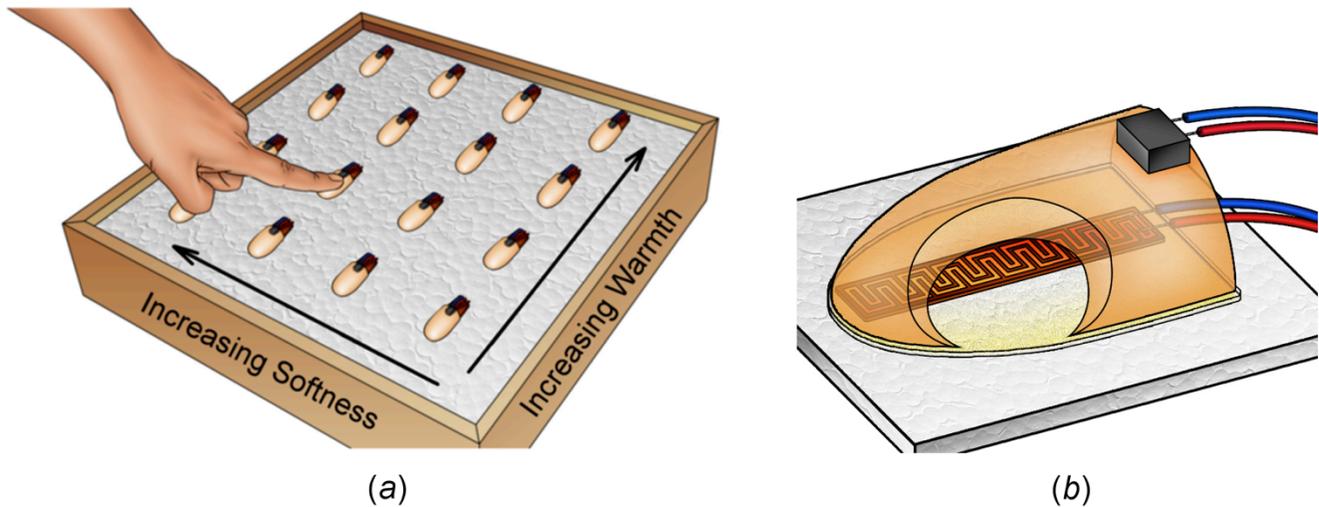

*(a)*          *(b)*

Fig. 2. Touch experiments on artificial skin samples. (*a*) Samples laid out in 4×4 array with columns arranged according to materials in increasing softness and rows arranged according to increasing warmth. (*b*) Cutaway view of an artificial fingertip skin sample consisting of an embedded heater and a temperature sensor mounted on the skin surface.

Each sample employed a hybrid on-off scheme and a proportional controller in order to reach the desired temperatures of 22, 27, 32 and 37°C. The algorithms were implemented in a microcontroller and a data acquisition system was integrated to it (Figure 3a). The initial skin surface temperature of the fingertip samples were measured with a temperature sensor (LM335, National Semiconductor, USA; accuracy ±0.5°C). The sensor was glued to the skin surface. The sensor was calibrated to room temperature before employing it on the circuit. The heater was made to operate in an on-and-off manner using metal–oxide–semiconductor field-effect transistor (MOSFET; IRF630, Fairchild Semiconductor, USA). A pull-up resistor was also used for matching the voltages of the microcontroller and the MOSFET. A microcontroller (Atmega 2560, Atmel, USA) coupled with an integrated development board (Arduino, Smart Projects, Italy) were used to maintain the temperature of the 16 fingertip samples to the desired temperatures. Shown in Figure 3b are representative readings collected from the data logger of a thermocouple-based multimeter (U1252A, Agilent, USA), which was attached to the surface of each material. The circuit design was then independently implemented on all the 16 fingertip samples.

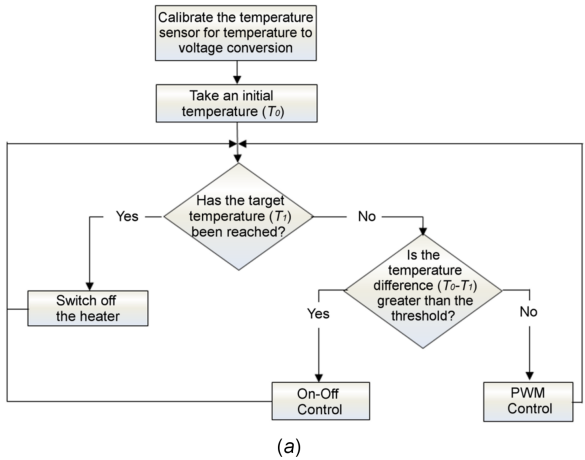

(a)

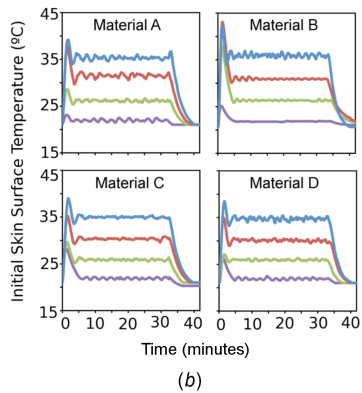

(b)

Fig. 3. Algorithm for the embedded heating and representative results. (a) The algorithm used for controlling the heating system to reach the desired temperatures of 22, 27, 32 and 37°C for each of the material sample. (b) Initial skin surface temperatures resulting from the heating system design.

## D. Thermal Characterization of the Materials

There will be a change in temperature at the contact surface when the human fingertip touches the skin material sample. In order to calculate that temperature, the thermal properties of the four materials were required. A thermal conductivity analyzer (TCi, C-Therm Technologies, Canada) directly measured thermal conductivity, $k$, and effusivity, $e$. The density, $\rho$, of each material was obtained from their respective data sheets. Specific heat, $c$, was then calculated as $c = e^2/k\rho$. Table II gives the thermal conductivity, density, and specific heat.

TABLE II. THERMAL PROPERTIES OF THE MATERIALS

| Material | Thermal Conductivity ($k$) (W/m·K) | Density ($\rho$) (kg/m$^3$) | Specific Heat ($c$) (J/kg·K) |
|---|---|---|---|
| A (Ecoflex) | 0.15 | 1,065 | 1,558 |
| B (Prochima) | 0.23 | 1,123 | 1,244 |
| C (Dragon Skin) | 0.16 | 1,081 | 1,435 |
| D (PDMS) | 0.12 | 1,040 | 1,610 |

*E. Initial Human Skin Temperatures*

The initial skin temperatures of the participants were measured as they entered the experiment room, and after 10, 20 and 30 minutes after they entered. We randomly selected 30 subjects (15 males, 15 females, all 18-25 years old) from those who participated in the skin sample experiments. The participants were made to sit in a relaxed position. The room temperature was maintained at 21°C. We measured the skin temperatures at the papillary whorl of the index fingertip of the participant's right hand using an infrared thermometer (model 62 Mini, Fluke, USA; accuracy ±1°C). The thermometer was maintained at a constant distance of 22 cm for all measurements. Upon entrance to the room, the mean temperature at the fingertip was 28.9°C (SD = 2.2°C). The succeeding mean temperatures were 29.0°C (SD = 2.0°C), 28.4°C (SD = 1.9°C) and 27.8°C (SD = 1.8°C), after 10, 20, and 30 minutes, respectively. For the subsequent calculations, the temperatures at the 10 minute mark were considered.

*F. Results*

The participants selected the samples shown in Figure 4a. Chance for selecting a sample is 6.25% (1 out of 16). For softness, 46% of the participants preferred material A while 32% preferred material B. For temperature, the 27°C skin surface temperature was selected by 45% of the participants while 31% selected the 32°C temperature. Notice that the participants' preferences clustered on material A with 27°C (n = 36), material A with 32°C (n = 26), and material B with 27°C temperatures (n = 28). By proportion, the preferred temperature results to 28.4°C. In summary, material A (Ecoflex OO-30) was the most preferred soft material. The preferred temperature can be set to 28.4°C based on the proportion of highly selected temperatures.

The majority's choice was consistent with the result of the heat transfer process that occurs when one touches a surface. Upon contact, the heat from the human fingertip was transferred to the surface of the artificial fingertip by conduction. The surface temperature at the interface of contact changed depending on the initial temperatures of the human and artificial fingertips and their thermal properties. We calculated the surface temperature of each material according to [34, 35]:

$$T_s = T_{artificial} + \frac{(T_{human} - T_{artificial})\sqrt{(k\rho c)_{human}}}{\sqrt{(k\rho c)_{human}} + \sqrt{(k\rho c)_{artificial}}} \qquad (1)$$

where $T_s$ is the resulting surface temperature upon contact of the human and artificial fingertips, $T_{human}$ is the initial temperature of the human fingertip, $T_{artificial}$ is the initial temperature of the artificial fingertip, $k$ is the thermal conductivity, $\rho$ is density, and $c$ is specific heat. The average initial temperature of the participants' fingertips was 29.0°C (§II.E). The thermal contact coefficient, $\sqrt{k\rho c}$, of the human fingerpad was estimated in [36] to be 1,181 J/m²s½K. The values for $k$, $\rho$, and $c$ for each of the artificial skin material were experimentally obtained (Table II). For Eqn. (1) to be valid, the Fourier number, $F_0$, of less than 5×10⁻² is the criterion to be satisfied to assume a semi-infinite model [37]. The Fourier number is

$$F_0 = \frac{\alpha t}{L_c^2} \qquad (2)$$

where contact time, *t*, was set to 5 seconds and the characteristic length, $L_c$, defined as material volume divided by the contact area, was calculated to be $4\times10^{-3}$ m. The thermal diffusivity of the material, $\alpha$, was calculated as $\alpha = k/\rho c$. The Fourier number for each material was found to be less than $2.1\times10^{-4}$. The $F_0$ criterion was satisfied. Thus, the use of Eqn. (1) is valid. The calculated surface temperatures at the contact interface are shown in Figure 4b. For the most preferred material and temperature combinations in Figure 4a, the calculated skin surface temperatures are 28.4, 28.3 and 29.9°C for material A with 27°C, material B with 27°C, and material A with 32°C, respectively. These temperatures are also closest to the average initial temperature of the participants' fingertips of 29.0°C (§II.E). Incidentally, the proportioned temperature was earlier calculated to be 28.4°C (§II.F).

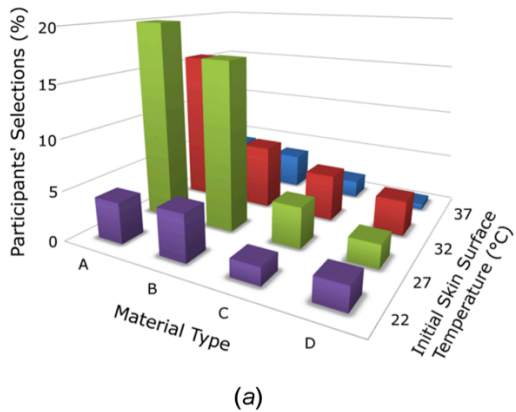

(*a*)

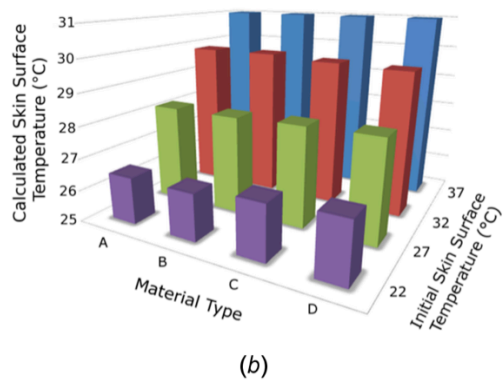

(*b*)

Fig. 4. Preferred warm and soft skin characteristics of the skin samples. (a) The selected material type and initial surface temperature of the artificial skins clustering at material A with 27°C temperature, material A with 32°C temperature, and material B with 27°C temperature. (b) The calculated temperatures at the contact interface when the human fingertip touches the skin samples.

## III. Lifelike Artificial Hand: Design, Construction, and Experiment

### A. Participant

A 35-year-old male subject participated in the study. The data from his dominant right hand were used to create a silicone rubber replica of a human hand. The subject's hand is 9 cm in breadth as measured across the ends of the metacarpal bones, and 19 cm in length as measured from the wrist crease to the tip of the middle finger. The subject was chosen because his hand's measurements were close to the 50[th] percentile of the anthropometric data in [38]. The experimental protocol for the indentation experiment was approved by the National University of Singapore's Institutional Review Board.

## B. Experimental Design

Two designs of artificial hands were fabricated in order to compare the softness of the human hand to the replicas made from silicone rubber. Here, softness is defined as the perceptual correlate of skin compliance, which is the amount of deformation caused by an applied force [39]. Both artificial hands were made from the data collected from the human subject's right hand. The first one made use of the skeleton structure while the second one did not. The various steps involved in fabricating the synthetic replica of the human subject's hand are shown in Figure 5.

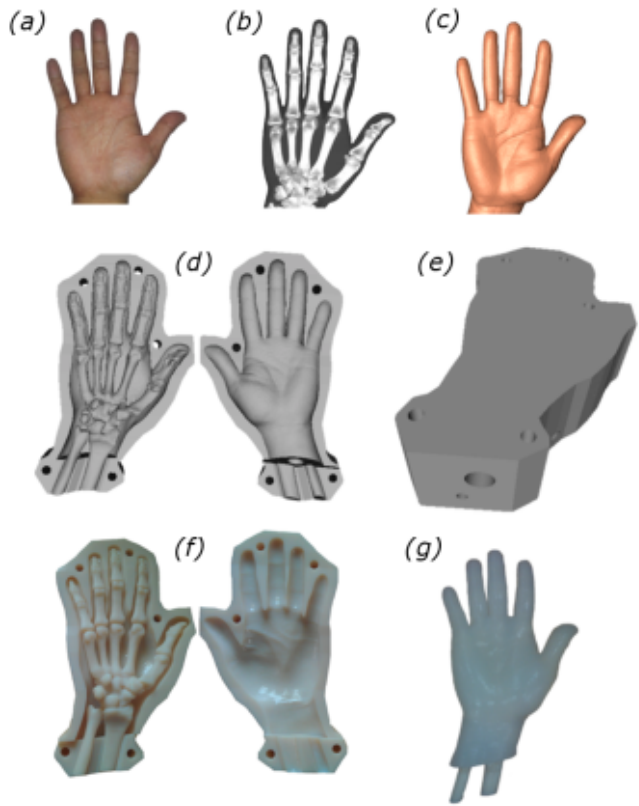

Fig. 5. The fabrication process to replicate a human hand. (a) The human subject's hand. (b) CT scan image showing the bone and the skin contours. (c) 3D surface reconstruction of the human hand. (d) The model of the two-part mould and the bones to construct a synthetic hand. (e) Assembled model of the mould. The liquid silicone rubber is poured through the large hole. (f) 3D printed model of the mould and the bones. (g) The completed replica of the human subject's hand obtained after the silicone rubber has cured.

Figure 5a shows the Computed Tomography (CT) scan of the subject's hand. A helical CT scanner (AquilionTM 64, Toshiba Medical Systems, Japan) used the following parameters during the scan: 120 kV, 150mA, 75 mA exposure, 500 ms exposure time and 0.5 mm slice thickness. The images were processed in the Vitrea 2 Imaging Software (Vital Images, USA) to produce a three-dimensional (3D) digital reconstruction of a human hand (Figure 5b and 5c). The data were post-processed with the 3-Matic software (Materialise, Belgium) to design a replica of the human hand's bones in stereolithography (STL) format. The same software was used to design the model of a two-part mould that would position the bones at accurate depths from the skin surface and have a provision for pouring liquid silicone rubber (Figure 5d and 5e). A 3D printing system (model 350, Objet Geometries, USA) with FullcureVeroWhite resin (Code 830, Objet Geometries, USA) was used for fabricating the STL models of the bones and the moulds (Figure 5f). The machine's printing resolutions are 600 dots per inch (dpi) in both the x and y axes and 1600 dpi in the z axis. The accuracy is up to 0.1 mm.

The material selected earlier (i.e. Ecoflex OO-30) was used as the synthetic skin material. The liquid silicone rubber was mixed and degassed in a vacuum chamber before being poured into the mould. After 24 hours of curing at room temperature of 21°C and relative humidity of 50-60%, the replica of the human hand with the skeleton was formed (Figure 5g). The fabrication process was repeated to obtain a second design of a hand with the same material, but without the presence of the skeleton. The use of the same two-part mould ensured equal outer dimensions of the two artificial hands. The one with the skeleton and the one without, are henceforth denoted as the artificial hand with bones (WB) and artificial hand with no bones (NB), respectively.

*C. Experiment setup*

The experimental setup used for the indentation tests is shown in Figure 6. The robotic arm (IRB-140, ABB Robotics, Sweden) has six degrees-of-freedom with a position repeatability of 0.03 mm. It is integrated with a force/torque sensor (ATI Industrial Automation, Apex, USA) with resolutions of 0.06 N in the x and y directions, and 0.125 N in the z direction. The indenter, a cylinder with a hemispherical tip of 2 mm diameter, was attached to the gripper on the robot's wrist to apply a constant force to the skin surface while the resulting displacement is measured (see Supplementary Material for the video of the experiments).

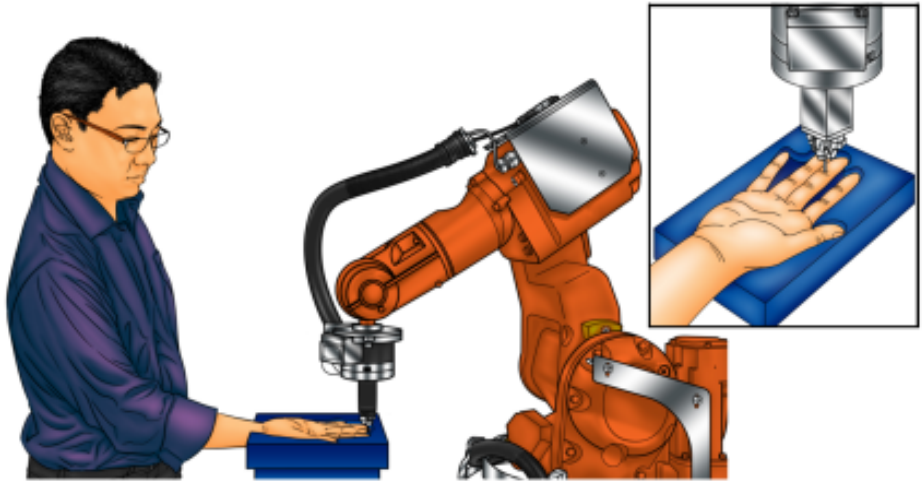

Fig. 6. Experimental setup. The robotic arm applied a constant indentation force on the hand. A force/torque sensor then measured the resultant displacement. The inset shows the hand as it rested on a clay mould to ensure a stable position with respect to the robot.

The hand being tested in the experiment was set in a supine position (Fig. 6 inset). The indentation tests were performed on predefined regions of the hands to create softness maps in order to compare the human and artificial hands. Those regions are displayed in Figure 7. A standardized method was used to identify the test boundaries, shown as ABCD and PQRS. We used the natural landmarks on the palmar side of the hand, like the papillary whorls and various creases, to define the boundaries.

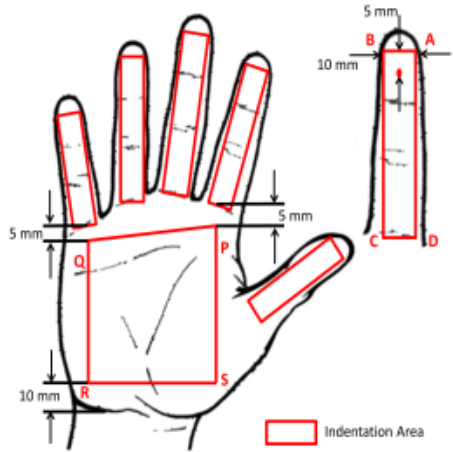

Fig.7. Standardized testing areas for indentation experiments on the human and artificial hands. Points ABCD and PQRS form the test boundaries that were programmed to the robotic arm

### D. Software Design

A top-level flowchart of the software code programmed in the ABB robot controller is shown in Figure 8. Programming was done using the RAPID programming language of the ABB robot system. To record the reference positions, the tip of the indenter was manually jogged to the corners of the boundaries defined in Figure 7. The robot was programmed to create a grid of test points within these predefined test boundaries. For the index, middle, and ring fingers, the program divided the test region into a 30×5 grid for a total of 150 test points. Due to their shorter length, the thumb and the little fingers were divided into a 25×5 grid with 125 test points each. The palm was divided into an 8×10 grid of 80 test points. Overall, 780 test points were available for the human and artificial hands and were sufficient to create a detailed softness map.

The final skin displacement was measured by using the displacement equation between two points $(x_1, y_1, z_1)$ and $(x_2, y_2, z_2)$ in three dimensional space, given by:

$$d = \sqrt{(x_2 - x_1)^2 + (y_2 - y_1)^2 + (z_2 - z_1)^2}$$

(3)

where $d$ is the displacement. There was negligible displacement seen on the $x$ or $y$ directions. The vertical displacement (i.e. $z$ axis) measured the amount of skin deformation produced by applying a force of 1 N.

### E. Data processing

Statistical analyses were performed using SPSS (version 21, IBM Corp., USA). For all analyses, statistical significance was set at a probability value of $p < 0.05$. A statistical comparison of the skin displacement data of the three hands was performed using one-way analysis of variance (ANOVA)[1]. A post-hoc multiple comparisons test was performed using the Tukey-Kramer Honestly Significant Difference (HSD) method.

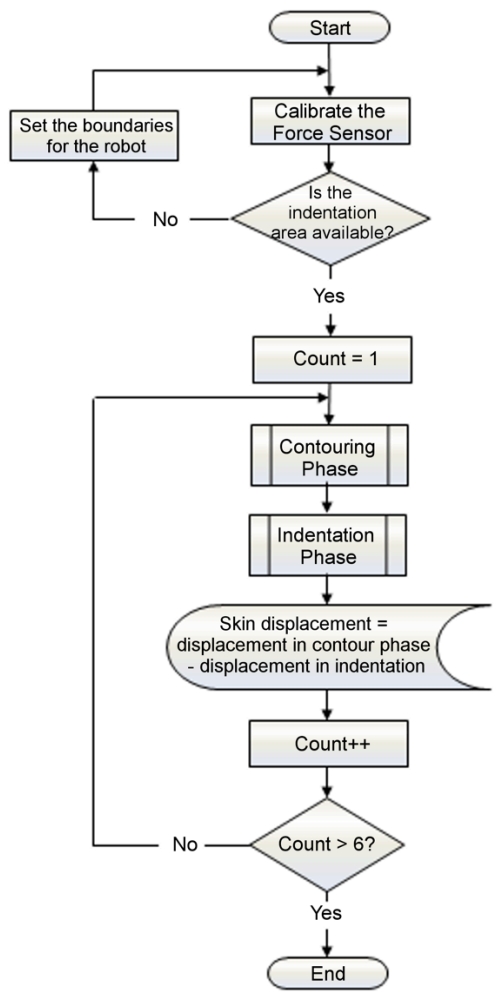

Fig. 8. A top-level description of the program flow for calibrating the sensor, setting the indentation boundaries, and for indenting the human and artificial hands.

### F. Skin softness map

The softness maps of the human and synthetic hands were derived from the skin displacement data and were plotted using contour plots in a software package (MATLAB, MathWorks, USA). A 2D coloured contour plot was created from the measured data and was superimposed on each type of hand (Figure 9). The raw data were processed with a moving average filter to smoothen the sudden gradient changes in the contour plots. The softness map of the human hand in Figure 9a indicated that the displacements were within 0.2 to 5.75 mm. Regions around the proximal interphalangeal (PIP) joints were observed to have the least displacement on all fingers. Figure 9b and 9c show the softness maps of the artificial hands with and without the presence of the skeleton. While the artificial hand with NB condition has displacements within the 2.5 to 7.5 mm, the displacements of the artificial hand having WB condition were within 0.5 to 6 mm and appeared to be closer to the human hand's skin compliance.

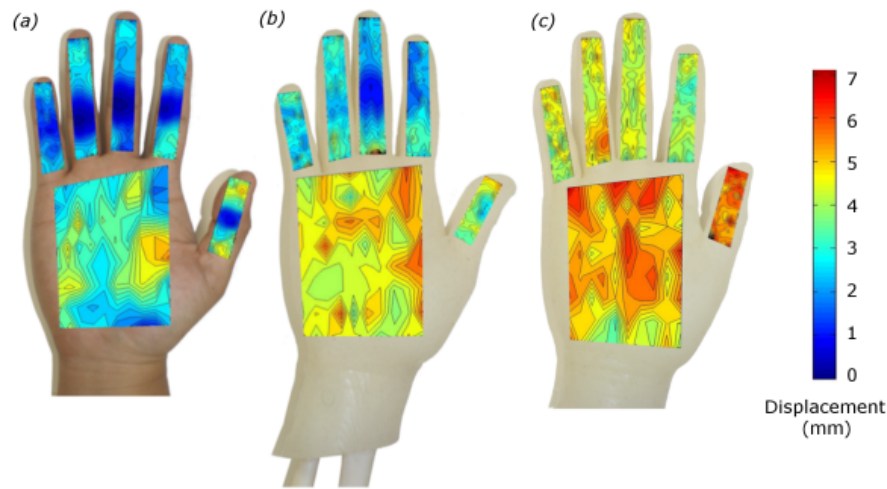

Fig. 9. Skin softness maps from the indentation tests with 1 N force for (a) human hand; (b) artificial hand with bone structure (WB); and (c) artificial hand with no bone structure (NB).

## G. Comparisons: human vs. artificial hands

Figure 10 compares the displacements of the human hand and the artificial hands. It consistently shows that the human hand has the least displacement, followed by the WB and NB artificial hands. A one-way ANOVA was performed on the displacements of the three groups (human, WB and NB hands) within each finger and the palm. The analyses yielded significant differences: thumb: $F_{(2,372)}$ = 222.8, $p < 0.001$; index: $F_{(2,447)}$ = 196.4, $p < 0.001$; middle: $F_{(2,447)}$ = 344.1, $p < 0.001$; ring: $F_{(2,447)}$ = 399.8, $p < 0.001$; little finger: $F_{(2,372)}$ = 375.7, $p < 0.001$; and the palm: $F_{(2,237)}$ = 128.1, $p < 0.001$. Post-hoc analyses showed that the displacements for human vs. WB hands, human vs. NB hands, and WB vs. NB hands differed significantly at $p < 0.001$. When the average displacements were compared, the artificial hand with bones differed from 14% to 47% against the human hand, while the artificial hand without bones differed from 54% to 143% against the human hand for every finger and the palm. Taken together, the artificial hand without bones was the softest among the three types of hands due to the absence of the bony structure. However, the displacements recorded for the artificial hand with bones were closer to the human hand.

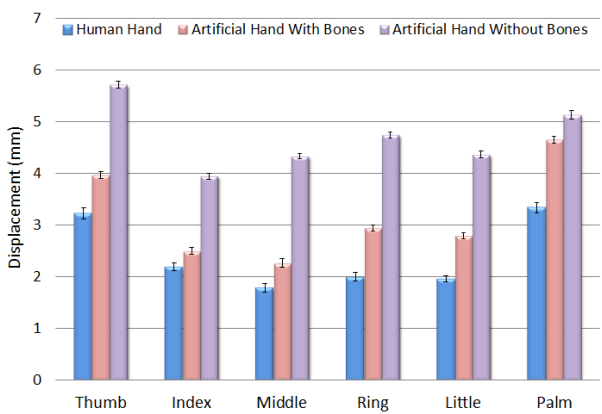

Fig. 10. Skin displacements at 1 N constant force (mean and SEM) for the human hand, artificial hand with bones (WB) and artificial hand with no bones (NB). Post-hoc analyses show significant differences ($p < 0.001$) for all the groups being compared for each of the fingers and the palm.

## H. Region-wise analysis

The geometrical structure of the bone and the external topography of the skin tissue are not uniform. Hence, a comparison was done between specific regions in each of the human and artificial fingers to investigate their effects on the displacement comparisons. For example, the first row of the 30×5 grid in the human index finger data was compared with the first row of the artificial index finger, and so on. In case of the palm, we divided the 8×10 grid into 2×2 regions for this comparison. A two-tailed paired t-test was performed between each set of human and artificial hand data. We matched the test hands in pairs as human vs. WB hand, and human vs. NB hand based on the region of measurement of the skin displacement. That is, displacement data from equivalent test regions of human and artificial hands were considered as one pair for the tests.

Figure 11 consolidates the paired t-test results by showing the individual test regions. The shaded areas in Figure 11a indicate a noticeable pattern of regions where no significant differences were found ($p > 0.05$) for the WB artificial hand and the human hand. This could be attributed to the presence of the thick volar ligaments at the PIP joints of the human hand, which can be 2 to 3 mm in thickness [40] (also see Figure 1). In comparison, the analysis of skin displacement data between the human and NB artificial hands (Figure 11b) shows few regions of similar skin compliance because of the missing bone structure.

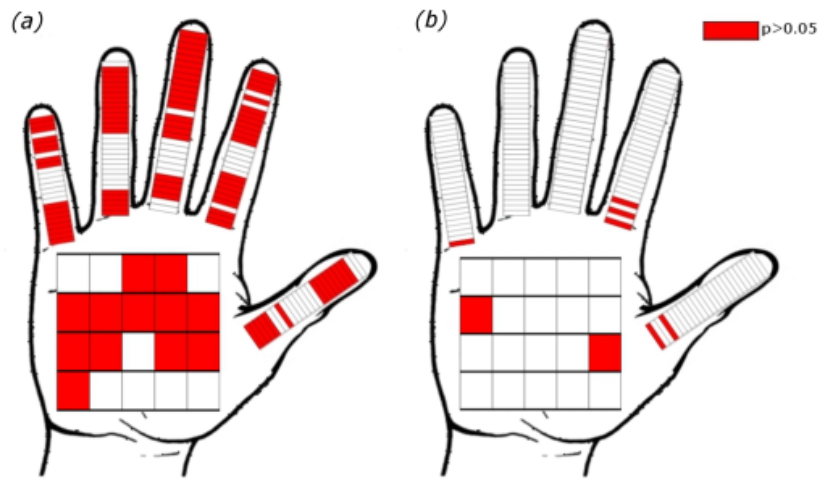

Fig. 11. Region-wise statistical comparison of human and artificial hands skin displacement data. (a) Human vs. artificial hand with bones and (b) human vs. artificial hand without bones. The shaded areas have *p*-values > 0.05, suggesting no significant difference between the artificial hand from the human hand.

## IV. Perception Experiments with Human and Artificial Hands

### A. Participants

A total of 28 healthy subjects (20 males, 8 females, all 17-26 years old) were recruited from the National University of Singapore. Participation was voluntary.

## B. Experimental Design

The participants were instructed to determine whether the hand that touches them is from a human hand or an artificial hand. We compared their responses after they were touched at the hairy part of the forearm by three types of hands (Figure 12a): a human hand, a soft and cold artificial hand, and a soft and warm artificial hand with the heating design. Both artificial hands had the bone structure (§III.H). The soft and cold artificial hand had no heating system and the skin surface temperature took the ambient temperature of 21°C. The soft and warm artificial hand was programmed with an initial skin surface temperature of 28.4°C (§II.F). The participants were naïve to the hypothesis of the experiment and to the types of hands that they will be touched with. During the experiment, the hands were kept out of the view of the participants (Figure 12b; also see Supplementary Video).

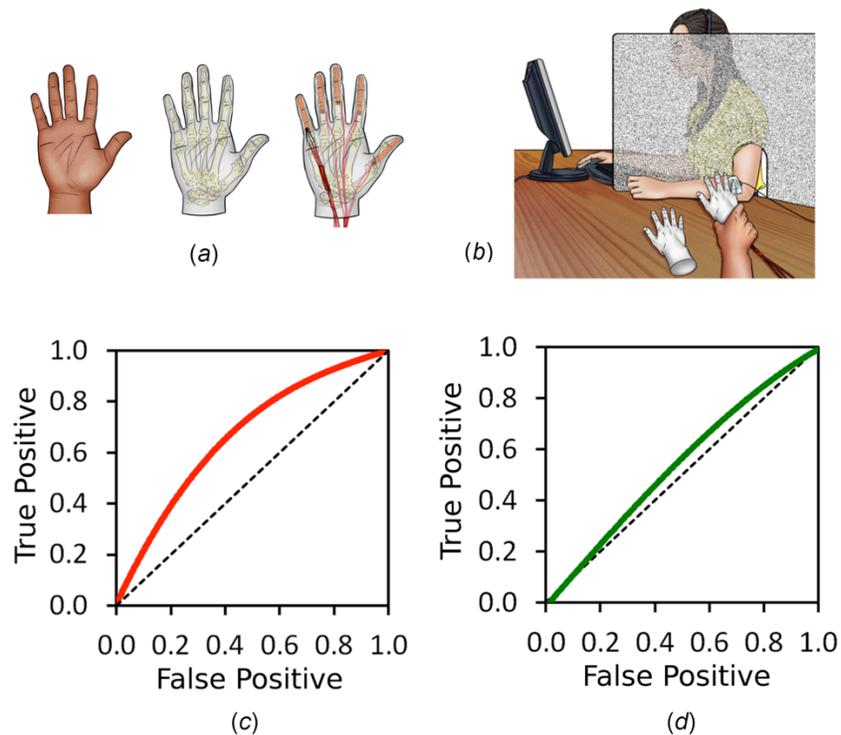

Fig. 12. Perception experiments with the human and artificial hands with bones. (a) From left to right: the human hand, soft and cold artificial hand, and soft and warm artificial hand. (b) Experimental set-up where the test hands touched the participants. (c) Receiving operating characteristics (ROC) curve of the soft and cold artificial hand and the human hand. (d) ROC curve of the soft and warm artificial hand and the human hand.

Detection theory [41] states that for experiments involving the participant's ability to tell two stimuli apart, two experiments may be distinguished. The first is *detection* where one of the two stimulus classes contains only the null stimulus (i.e. yes or no task). The second one is *recognition* where neither stimulus is null (i.e. rating experiments on a scale). Rating experiments give more information than in a "yes-no" task. For the current paper, we asked participants to respond on a 5-point scale where 1 denotes that they are certain that the touch is from an artificial hand and 5 denotes that they are certain that the touch is from a human hand. With a computer mouse, participants entered their responses according to a 5-point scale. A force sensor (FingerTPS,

Pressure Profile System, USA) was positioned at the participant's forearm to help the experimenter regulate the amount of contact force. The applied force was about 1 N. The three hand types were equally presented in a pseudorandom order for a total of 27 trials. The first 6 trials were discarded to account for practice effects. A total of 588 responses (28 subjects, 21 trials each) were obtained. This data set was sufficient as it was suggested that at least 100 observations are needed in receiver operating characteristics (ROC) curve analysis to draw a meaningful conclusion [41].

*C.  Data processing*

We used the ROC curve analysis [41] to relate the proportion of correctly recognized touches by the warm or cold artificial hand (true positive) to the proportion of mistakenly recognized touch from the human hand (false positive). The differences between the two Areas Under the Curves (AUC) were compared with the method of Hanley and McNeill [42]. In this method, the p-value was determined as the probability that the observed sample AUC is found when the true population AUC is 0.5 and that the null hypothesis assumes that the area is 0.5. The probability selected as significant was $p < 0.05$.

*D.  Results*

For each type of hand that was used to touch the subjects, Table III shows how the subjects recognized them from 1 (very certain to be prosthetic hand) to 5 (very certain to be human hand). These data were further analyzed using ROC curves. A result that cannot discriminate between correctly and mistakenly recognized touch corresponds to an ROC curve that is coincident to a diagonal line and occupies an AUC of 0.5. In other words, an AUC closer to 0.5 suggests that the touch from an artificial hand could not be recognized as a touch from a human hand. Figure 12c shows that the touch from the soft and cold artificial hand can be recognized from the human hand (AUC = 0.697, $p < 0.0001$).

In contrast, Figure 12d suggests that the participants had difficulty in recognizing the soft and warm artificial hand from the human hand (AUC = 0.562, $p = 0.0611$). The p-value for the cold hand was found to be less than 0.05 indicating that the AUC is significantly different from an area of 0.5. For the warm artificial hand, the p-value was found to be greater than 0.05 suggesting that the touch from this artificial hand could pass as if it were touch from a human hand.

TABLE III FREQUENCY OF EACH RESPONSE FOR EACH STIMULUS (TOUCH RECOGNITION)

| Hand type | Very Certain to be Prosthetic (1) | Certain to be Prosthetic (2) | Neutral (3) | Certain to be Human (4) | Very Certain to be Human (5) | Sum |
|---|---|---|---|---|---|---|
| Human | 21 | 35 | 24 | 65 | 51 | 196 |
| Cold | 49 | 49 | 33 | 46 | 19 | 196 |
| Warm | 20 | 34 | 52 | 45 | 45 | 196 |

## V. Discussion and Conclusion

For over a century, the prosthetic hook [43] has retained its basic design. It has been a robust and reliable terminal tool for simple grasping tasks. However, the rejection rate for the hook-type prosthesis has been high at 50% among upper limb amputees [44].

Continuing advancements from several research domains are now converging towards the development of more intelligent prosthetic devices. Prosthetic hands can perform dexterous grasping (e.g. [21, 45, 46]); it can be controlled by amputees according to their motor commands [6, 47, 48]; an object's size or shape can now be felt by an amputee [7, 8]; and through experiments on the rubber-hand illusion, it has also been demonstrated that an artificial hand can be felt as if it were a part of the body [12, 16, 49]. All these are giving upper limb amputees hope to regain what has been lost.

Considering the technological trends above, it can be expected that prosthetic hands will touch and be touched by others during social interactions. To our knowledge, the current paper is the first to demonstrate that a warm and soft silicone artificial hand can be used to create an illusion of the sense of human touch. This was accomplished through a three-step process. First, participants were asked to select the samples that they felt were similar to human skin from artificial skin samples that were laid out in a 4×4 array. Majority selected a silicone material with a Shore durometer value of 30 at the OO scale and the selected temperatures clustered at 27 and 32°C in an ambient room temperature of 21°C. When proportioned, the selected temperature was calculated to be 28.4°C. Our results suggest that a soft rubber material with a warm skin surface temperature is critical for the artificial skin to be perceived as lifelike. Rubber materials are known to be poor conductors of heat. The thermal conductivities of the 4 materials tested herein were low with values from 0.12 to 0.23 W/m·K. However, it would be possible to mimic the warmth of the human hand's skin tissue by controlling the embedded heater's power supply by adjusting the heat that will be transmitted to the surface of a rubber material [27, 50]. A more lifelike artificial skin is the one having a surface temperature that is similar to the initial surface temperature of the human skin upon contact. This means that when a person touches a sample, it tends to feel more lifelike when the person does not experience a significant difference on the warming or cooling sensation on his/her finger at the moment of contact. The contact temperature will naturally rise and approach the human body temperature due to the elimination of the natural air convection from the exposed finger.

Second, we described a method to replicate the geometric features of a human subject's hand and the softness of the hand's skin tissue. Through CT scan, computer-aided design, 3D printing and silicone moulding technologies, we found that the softness of the human hand's skin tissues can be mimicked by replicating the surface topology of the human hand and the geometries of its skeleton. The soft material selected in the 4×4 array experiment was used as a substitute for the human skin tissue. Due to the absence of ligaments and tendons like those in human hands, there were regions on the fingers and the palm that were softer than the human hand. Overall, the DIP and PIP regions were not significantly different from the human hand ($p > 0.05$). We used these regions of the artificial hand to touch the forearm of the participants. To integrate other mechatronic components for a more intelligent prosthetic hand, there is still substantial space that is available for mechanisms and embedded electronics at the skeleton and at the dorsal part of the hand. From our initial design, only 5-7 mm of artificial skin thickness are needed to be allocated for embedded heaters and for the soft material to replicate the human hand's skin compliance.

Finally, we combined the previous results to design the artificial hands for the perception experiments. Subjects who were naïve about the experimental objectives were touched with a human hand and two artificial hands. One artificial hand was soft but it felt cold because the skin surface temperature was similar to that of the room temperature of 21℃. The other artificial hand felt warm due to the embedded heaters, which raised the skin surface temperature to 28.4℃. With the participants' field of view being restricted, they were asked to recognize whether they were touched by a human or an artificial hand. The subjects were touched at the hairy part of the forearm where a subclass of unmyelinated afferents (C-tactile, CT) are known to be present [51]. The CT afferent has been implicated in the coding of pleasant tactile sensations. ROC curve analysis suggests that an illusion of human touch can be created from an artificial hand with embedded heaters on soft synthetic skin and a skeleton structure (AUC = 0.562; p = 0.0611).

An early study suggested that friction and the surface topography of a material are significant factors when the human finger pad makes a sideward movement on a surface, which can further influence the sensation of pleasant touch [52]. The current paper only investigated simple touches of 1 N normal force and not the sliding touch movements that are similar to a caress. Such movements would entail controlled lateral movements of 1-10 cm/s [53]. Likewise, the fingerprint ridges and the sweat from the skin pores have been shown to contribute to the contact mechanics of sliding movements [54]. In future studies, it would be interesting to investigate the effects of artificial fingerprint ridges on the pleasantness of the touch through controlled lateral movements similar to a caress.

More advanced touching movements, like caress and handshakes, have social, emotional, and cultural ramifications. The present work focused only on simple mechanical touch. Despite its seeming simplicity, social touching for artificial hands has been a neglected research area. From the experiments described herein, we found that softness and warmth as important features. Future work can look into the other design variables that can make social touches (e.g. handshake or caress on the face, the hands, and arms) non-discriminable even through an artificial hand.

Interpersonal touch induces strong and reliable changes in autonomic activity (e.g. skin conductance, pulse and respiration) between the interacting partners [55]. A possible extension to the current work is to investigate the emotional valence of the touch from a lifelike artificial hand. That is, we might be able to see effects of this novel type of prosthetic hand for affective touches. There have been related works on determining the emotional state of participants by touch from haptic interfaces. Some of the earlier works have explored that idea with a haptic jacket [56] or a haptic sleeve like in our earlier work [57]. However, there have been none for prosthetic or robotic hands. In the Affective Teletouch Technology that we proposed [57], we asked participants to watch a sad movie while the spouse touched the arm of the participant. In the other experimental condition, we used a haptic sleeve to provide vibratory and warmth stimuli on the subject's arm. A similar experimental design could be used for the soft and warm artificial hand as described herein.

For practical implementation later on, it would be helpful to consider a layered skin structure that addresses the multiple requirements to have soft features for social touching [28], for skin compliance and skin conformance for tactile sensing purposes [58] as well as the need to protect the underlying structures through a tough skin layer [50]. In an earlier work [29], we investigated how to design a soft skin for social touching interactions, which also considers other requirements for wear, puncture, and tear. For the condition of having a softer internal material and a stiffer external material with a 0.8 mm thin layer, we found that the results of a layered structure will only have a difference in displacement of about 7% as compared to a structure that is homogeneous. These previous findings suggest that it would be possible to achieve both skin softness as well as toughness by varying the properties of the skin layers.

The present results are important because they provide an early evidence that an illusion of human touch can be created by a warm and soft silicone artificial hand to the person being touched. When applied to prosthetic hands, these findings have a potential to help prosthesis users cope with the functional and psychosocial effects of losing a part of their body.

### Acknowledgment

This work was supported in part by the Research and Graduate Studies Office of the College of Engineering, Qatar University and the Academic Research Fund of the National University of Singapore. The authors are grateful to Jaclyn Ting Lim for the illustrations.

**John-John Cabibihan** received his PhD in the area of biomedical robotics from the Scuola Superiore Sant'Anna, Pisa, Italy in 2007. Concurrent with his PhD studies, he was awarded an international scholarship grant in 2004 by the Ecole Normale Supérieure de Cachan, France. Therein, he spent one year at the Laboratoire de Mécanique et Technologie. From 2008 to 2013, he was an Assistant Professor at the Electrical and Computer Engineering Department of the National University of Singapore, where he also served as the Deputy Director of the Social Robotics Laboratory and an Affiliate Faculty Member at the Singapore Institute of Neurotechnologies (SiNAPSE). He is presently an Assistant Professor at the Mechanical and Industrial Engineering Department of Qatar University. He serves at the Editorial Boards of the International Journal of Social Robotics, International Journal of Advanced Robotics Systems, Frontiers in Bionics and Biomimetics, and Computational Cognitive Science. He was a past Chair of the IEEE Systems, Man and Cybernetics Society, Singapore Chapter (terms: 2011 and 2012). He was also the General Chair of the 6th IEEE International Conference on Cybernetics and Intelligent Systems (2013), Program Chair of the 4th International Conference on Social Robotics (2012), and Program Co-Chair of the 2nd International Conference on Social Robotics (2010). Over the years, his work has been focused on assistive and social robotics for the therapy of children with autism, lifelike prosthetics, bio-inspired tactile sensing, and human-robotic touch and gestures.

**Deepak Joshi** received the B. Tech in Instrumentation engineering and M. Tech in Instrumentation and control engineering from India in 2004 and 2006, respectively. He received PhD degree in biomedical engineering from Indian Institute of Technology (IIT) Delhi. He was a visiting scholar at Institute of Neuroscience (ION), Newcastle University and research engineer at national university of Singapore (NUS), before joining Graphic Era University, Dehradun India. Currently, he is on leave from the university to pursue his postdoctoral research at University of Oregon, USA. His research area includes neural-machine interface, machine learning, biomedical instrumentation, and signal processing.

**Yeshwin Mysore Srinivasa** received his Master of Science degree in Electrical Engineering from the National University of Singapore, with a specialization in Automation and Control Engineering. He is presently working as a Research and Development Engineer at Makino Asia Pte Ltd. His research interests include industrial robotics and high-speed motion control systems.

**Mark Aaron Chan** is a Research Engineer at GE Global Research, Germany. He received his PhD in Mechanical Engineering in 2010 from National University of Singapore. Subsequently, he joined the Cryospheric Sciences Laboratory at NASA GSFC as a postdoctoral fellow. His main research interests are boiling heat transfer and thermal management of electronics.

**Arrchana Muruganantham** received the B.Eng degree in Electrical Engineering with Honors from the National University of Singapore, Singapore in 2012. Since then, she has been pursuing her PhD degree in Electrical Engineering with the Department of Electrical and Computer Engineering from the same university. Her current research interests include Computational Intelligence and Robotics, in general and Evolutionary Computation, in particular.